\documentclass[aps,prl,twocolumn,floatfix,english,showpacs,10pt,superscriptaddress]{revtex4-2}%
\usepackage{graphicx}
\usepackage{booktabs}
\usepackage{amsmath}
\usepackage{physics}
\usepackage{amssymb}
\usepackage{colordvi}
\usepackage{verbatim}
\usepackage{xcolor}
\usepackage{mathrsfs}
\usepackage{epsfig}
\usepackage{lipsum}
\usepackage{amsfonts}
\usepackage{makecell}
\usepackage{esint}
\usepackage{bm}
\usepackage{float}

\usepackage[most]{tcolorbox}

\usepackage[unicode=true, breaklinks=false, pdfborder={0 0 1}, backref=false,
colorlinks=true, linkcolor=blue, urlcolor=blue, citecolor=blue]{hyperref}%
\providecommand{\U}[1]{\protect\rule{.1in}{.1in}}

\providecommand{\U}[1]{\protect\rule{.1in}{.1in}}
\setcitestyle{numbers,square}

\begin{document}

\title{Fractional Magnonic Frequency Combs}

\author{Qian-Nan Huang}
\affiliation{School of Physics, Huazhong University of Science and Technology, Wuhan 430074, China}

\author{Zhiping Xue}
\affiliation{School of Physics, Huazhong University of Science and Technology, Wuhan 430074, China}

\author{Xudong Wang}
\affiliation{School of Physics, State Key Laboratory of Crystal Materials, Shandong University, Jinan 250100, China}

\author{Yanmeng Lei}
\affiliation{School of Physics, Huazhong University of Science and Technology, Wuhan 430074, China}

\author{Lihui Bai}
\email{lhbai@sdu.edu.cn}
\affiliation{School of Physics, State Key Laboratory of Crystal Materials, Shandong University, Jinan 250100, China}

\author{Ke Xia} 
\email{kexia@seu.edu.cn}
\affiliation{School of Physics, Southeast University, Jiangsu 211189, China}

\author{Gerrit E. W. Bauer}
\email{bauer.gerrit.ernst.wilhelm.d8@tohoku.ac.jp}
\affiliation{WPI-AIMR and Institute for Materials Research and CSIS, Tohoku University, Sendai 980-8577, Japan}
 
\author{Tao Yu}
\email{taoyuphy@hust.edu.cn}
\affiliation{School of Physics, Huazhong University of Science and Technology, Wuhan 430074, China}

\date{\today }

\begin{abstract}

Magnonic frequency combs (MFCs) are spectacular phenomena in microwave-driven high-quality magnets. Like the equally spaced prongs in a comb, conventional \textit{integer} MFCs are sharp resonances with an equal and constant frequency difference. Here we report \textit{fractional} MFCs in a high‑quality magnetic sphere that emerges when adding a low‑power, precisely detuned microwave to the main drive that compresses the frequency spacings to a rational fraction of the original comb, generating high‑density spectral grids with hundreds of lines. The theoretical analysis finds that parametric three‑magnon scattering is the dominant non-linear process that reproduces the observation well. This mechanism is unique to magnets: it does not exist in an optomechanical system, where the Kerr and optical nonlinearities govern comb formation at a much higher power input. Since our platform operates as a frequency ``vernier caliper" with much higher sensitivity than integer MFCs, it has application potential in precision metrology.

\end{abstract}

\maketitle

Optical spectra comprising a series of discrete, equidistant spectral lines are referred to as frequency combs~\cite{Udem2002OpticalMetrology,Cundiff2003FemtosecondComb,Hansch2006Nobel,DelHaye2007comb,Fortier2019comb,Hillbrand2020synchronization}.  They are extensively deployed in cutting-edge fields ranging from atomic clocks~\cite{papp2014microresonator,Ludlow2015OpticalAtomicClocks}  over precision metrology~\cite{RevModPhys781279,trocha2018ultrafast,wang2020long} to optical communication~\cite{2017microresonator,tan2018multichannel,2020ultradense}. 
Magnons---the quanta of spin waves---in high-quality magnetic insulators combine ultralow magnetic damping and exceptional frequency tunability at room temperature~\cite{serga2010yig,magnonics1,bauer2012spin,chumak2015magnon,magnonics3,demidov2017magnetization,Bistability057202,manchon2019current,Chiral027203,Coherent243601,Yu2024,2025coherent}. Translating the concept of frequency combs from photons into the magnon domain~\cite{liu2018magnon,Skyrmion037202,Spin2022,Skyrmion202101245,xiong2023magnonic,053708Generation,Magnonic243601,RaoPhysRevLett,Design174412,PhysRevA023507,Liu2024,2024enhancement,Nonreciprocal2025,Mechanically063703,LanCoherent2025,Heins2026} promises similar leaps in coherent information processing by leveraging the exceptional tunability by and strong coupling to microwaves. Over the past several years, magnonic frequency combs (MFCs) have been addressed through diverse nonlinear pathways: the magnon Kerr effect (four-wave mixing) in high-Q spheres and microstructures~\cite{liu2018magnon,RaoPhysRevLett,2024enhancement,Nonreciprocal2025,053708Generation}, magnomechanical coupling to deformation‑potential phonons~\cite{xiong2023magnonic,053708Generation,Mechanically063703,Magnonic243601}, four-magnon scattering in confined waveguides~\citep{Spin2022}, and, in textured films, three-magnon processes lock the neighboring tooth to a texture mode~\cite{Skyrmion037202,Skyrmion202101245,Design174412,LanCoherent2025,Heins2026}. Yet a shared bottleneck persists: in the experimental implementations, the comb step is rigidly tied to the characteristic frequency of an underlying hybrid mode or the periodic drive, so the spectral lattice is coarse, typically fewer than $\sim$30 resolved teeth, and its grid cannot be programmed independently of the device characteristic frequencies.

\begin{figure}
    \centering
    \includegraphics[width=1\linewidth]{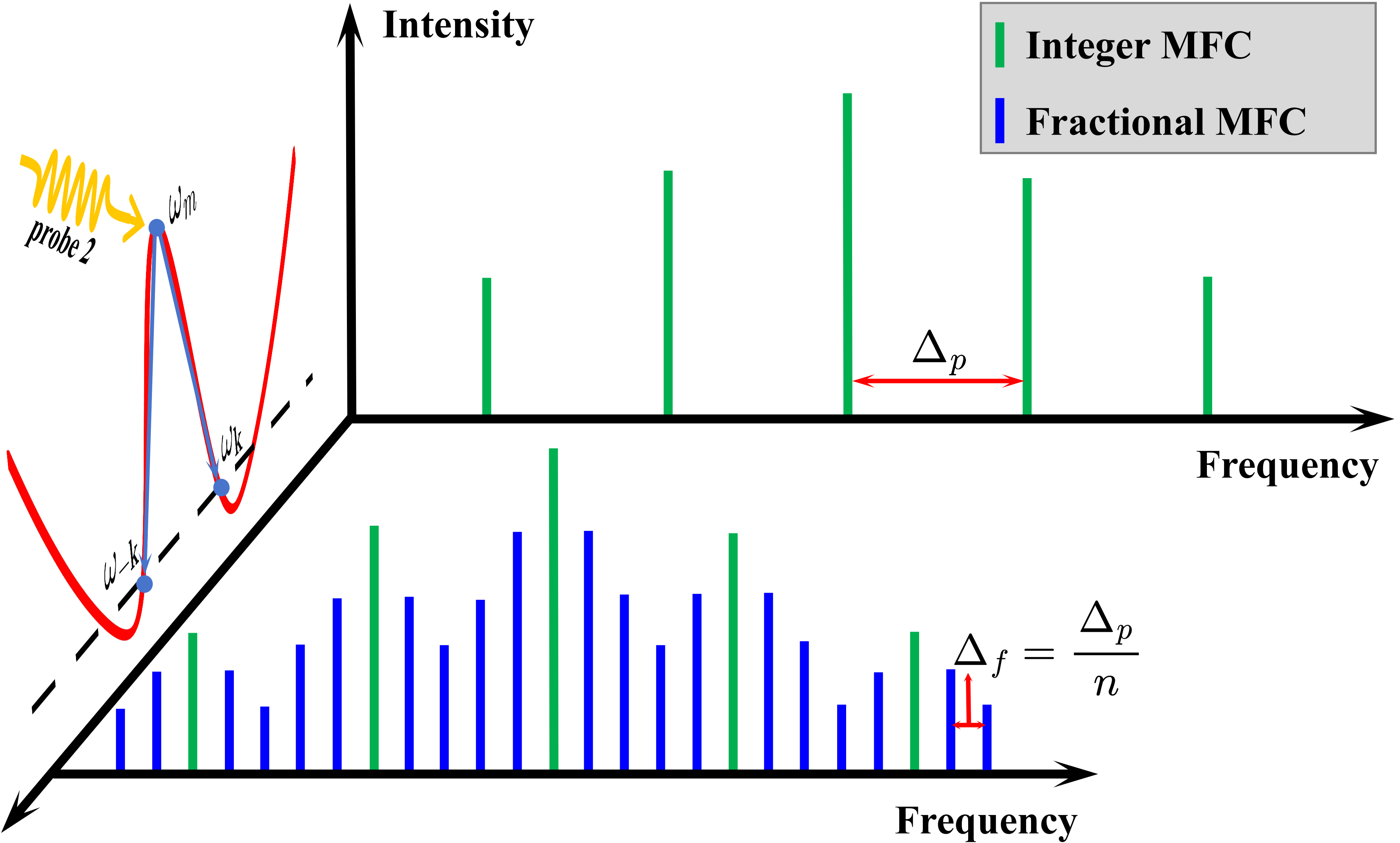}
    \caption{Principle of fractional MFC via parametric three-magnon splitting. An integer MFC, driven by two microwave tones with detuning $\Delta_p$, yields only a limited number of spectral lines. The introduction of a third drive (probe 2) with a finely tuned detuning $\Delta_f=\Delta_p/n$ ($n$ is an integer) triggers a massive nonlinear cascade via the parametric three-magnon process. In this mechanism, the Kittel magnon ($\omega_m$) splits into two degenerate magnon modes ($\omega_{\pm\bf k}=\omega_m/2$), generating a fractional MFC with a significantly enhanced number of teeth and superior spectral resolution.}
    \label{Fig1}
\end{figure}

Here, we overcome this limitation by reporting the experimental discovery of a new kind of MFC, i.e., a fractional MFC. By introducing a low-power, precisely detuned third microwave to the main drive, we decouple the comb step from intrinsic eigenfrequencies, compressing it to a rational fraction of the original interval and generating high-density spectral grids exceeding 250 lines. While the concept of fractional-order sideband generation was proposed theoretically in optomechanical systems relying on optical nonlinearity~\cite{liu2020fraction,liu2021tunable,liu2025multiple}, we demonstrate that three-magnon scattering---rather than Kerr~\cite{2023nonreciprocal} or magnetostrictive effects~\cite{wu2025fraction,zhou2026magnon}---dominates the generation process in the magnets. The principle of generating a fractional MFC by the parametric three-magnon process~\cite{PumpL180410,Magnon166703,Huang2026} is illustrated in Fig.~\ref{Fig1}. This mechanism enables compression of the comb spacing to $\sim 1/20$ of the original interval with a power $\sim 3~\mu$W, three orders of magnitude lower than that used in optomechanics~\cite{liu2020fraction}. Furthermore, we demonstrate that this architecture functions as a frequency-domain ``vernier caliper," amplifying frequency resolutions by an order of magnitude compared to that of conventional integer MFCs. Beyond demonstrating a high-density spectral grid, this work demonstrates that three-magnon scattering acts as the dominant, engineerable nonlinearity in high‑$Q$, large‑volume magnetic spheres, opening a previously underused degree of freedom for programmable microwave responses.

\begin{figure}
    \centering
\includegraphics[width=1.0\linewidth]{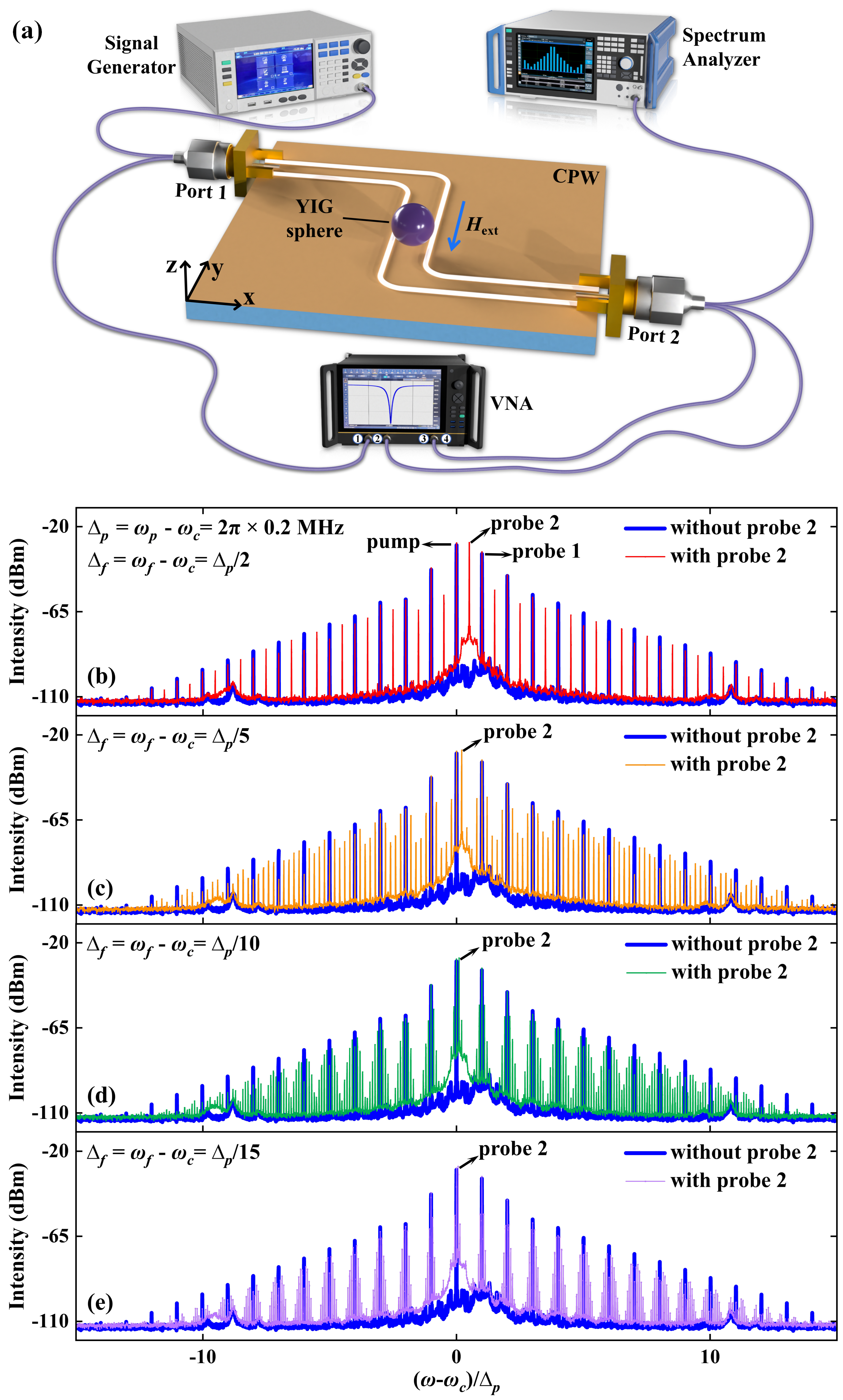}
\caption{Experimental realization of fractional MFC. (a) A YIG sphere at the midpoint of the CPW central strip is subject to an external static magnetic field $-H_{\rm ext}\hat{\bf y}$. The system is driven by a strong pump microwave and two probe microwaves, generated by the signal generator and VNA, while the microwave spectra are monitored by the spectrum analyzer. (b)-(e) Comparison of microwave spectra without and with ``probe 2" microwaves at four different detuning $\Delta_f = \Delta_p/2$, $\Delta_p/5$, $\Delta_p/10$,  and  $\Delta_p/15$, respectively. The blue curves represent the generated integer MFC when only the pump and probe 1 microwaves are applied; the curves in the other colors denote the generated fractional MFC when subject to all three microwave drives.}
\label{Fig2}
\end{figure}

Figure~\ref{Fig2}(a) illustrates the experimental configuration. A coplanar waveguide (CPW) is designed on a copper substrate (lateral dimension 25~mm$\times$25~mm, thickness 0.03~mm, and standard impedance 50~$\Omega$) and features a central strip of width 0.8~mm. An yttrium iron garnet (YIG) sphere with a diameter of $1$~mm is mounted at the center of this strip. The assembly is placed horizontally at the center of an electromagnet, applying a static magnetic field $-H_{\rm ext}\hat{\bf y}$ within the CPW plane and aligns parallel to the central strip.  We characterize the microwave transmission spectra $S_{21}$ across the YIG sphere under a static bias magnetic field $\mu_{\rm 0} H_{\rm ext}=104.308$~mT to determine the ferromagnetic resonance (FMR) frequency  $\omega_{m}=2\pi\times 3.0000$~GHz~\cite{Yao2019,YuChiral2020,Rameshti2022,Walker1957Magnetostatic, Dillon1957YIG,Gloppe2019Resonant,serga2010yig}. As shown in Fig.~\ref{Fig2}(a), three microwave drives are injected via a microwave signal generator and a vector network analyzer (VNA), in which the drive from the signal generator serves as the pump ($\omega_c$), while the VNA injects two additional probes detuned from the FMR frequency: probe 1 ($\omega_p$) and probe 2 ($\omega_f$). To generate the MFC, the frequency detunings are modulated such that  $\Delta_{p}=\omega_{p}-\omega_{c}$ and $\Delta_{f}=\omega_{f}-\omega_{c}=\Delta_{p}/n$, where $n$ is an integer defining the rational division ratio. The resulting frequency-comb spectra are gathered using a spectrum analyzer connected to the CPW output port.

A conventional integer MFC is generated when driven solely by the pump and probe 1 microwaves (i.e., without probe 2). To this end, the pump frequency is set to match the FMR frequency, i.e., $\omega_{c}/(2\pi)=\omega_{m}/(2\pi)=3.0000$~GHz, with a power of $P_{c}=-10$~dBm [see Supplementary Material (SM)~\cite{supplement} Sec.~I for the power dependence of integer MFC]. Simultaneously, the probe 1 is tuned to  $\omega_p/(2\pi)=3.0002$~GHz, yielding a detuning of $\Delta_{p}/(2\pi)=(\omega_{p}-\omega_{c})/(2\pi)=0.2$~MHz relative to the pump, with its power adjusted to $P_{p}=-15$~dBm (see SM~\cite{supplement} Sec.~I for the power dependence). Under these conditions, we observe an integer MFC characterized by a tooth spacing of $\Delta_{p}/(2\pi)$ and fewer than 30 lines, as shown by the ``without probe 2" traces in Fig.~\ref{Fig2}(b)-(c).

We now introduce the probe 2 drive with a low power of $P_{f}=-25$~dBm (3.2~$\mu$W) to the CPW, and adjust the detuning relative to the pump microwave $\Delta_{f}=\Delta_{p}/n$, where the integer $n$ is systematically varied in our experiments. 
Figures~\ref{Fig2}(b)-(e) illustrate the emergence of the fractional MFC and its evolution with $n$, with a direct comparison for $n=2$, $5$, $10$, and $15$ shown in Figs.~\ref{Fig2}(b)-(e). The blue curves depict the conventional integer MFC (28 teeth, spacing $\Delta_p$), while the curves in the other colors show the fractional MFC triggered by probe 2. Remarkably, this weak probe suffices to initiate a massive nonlinear cascade, multiplying the number of teeth and drastically enhancing spectral resolution.
Indeed, by tuning the frequency detuning to $\Delta_{f}=\Delta_{p}/2=2\pi\times 0.1$~MHz in Fig.~\ref{Fig2}(b), we observe the new $\pm l/2$-order fractional sidebands alternating with the pre-existing teeth, effectively doubling the total number of comb teeth to $56$.
When the integer $n$ increases to $5$, as shown in Fig.~\ref{Fig2}(c), the detuning of probe 2 microwave becomes $\Delta_f = \Delta_p/5$ and four new fractional-order sidebands are uniformly inserted between each pair of adjacent integer comb teeth, elevating the comb count to $130$ and increasing the frequency resolution to five times that of the original spectra. As the detuning is further decreased to $\Delta_f = \Delta_p/10$ in Fig.~\ref{Fig2}(d) and $\Delta_f = \Delta_p/15$ in Fig.~\ref{Fig2}(e), the integer frequency sidebands are surrounded by high-density fractional frequency teeth, reducing the frequency intervals to $20$~kHz and beyond, with the total number of teeth reaching $223$ and $256$, respectively.

The continuous tunability of the fractional MFC reveals the evolution from an integer MFC to a fractional MFC comprising three distinct spectral families: (i) integer-order frequency sidebands at $\Omega_{\rm I} = \omega_c\pm j\Delta_p$; (ii) fractional-order frequency sidebands distributed symmetrically around the pump frequency at 
$\Omega_{\rm II} = \omega_c \pm l\Delta_f = \omega_c \pm \frac{l}{n}\Delta_p$; and (iii) the sum- and difference-frequency fractional sidebands spaced at intervals of $\Delta_p/n$ around other integer-order sidebands at 
$\Omega_{\rm III}=\omega_c\pm j\Delta_p \pm l\Delta_f=\omega_c \pm \left(j \pm \frac{l}{n}\right)\Delta_p$, where $j$ and $l$ are positive integers. This rational fraction $1/n$ acts as a magnification factor for frequency resolution. The number of comb teeth scales linearly with the fraction control parameter $n$. Under ideal conditions, the total comb count follows $N_{\rm total} = n \times N_{\rm int}$, where $N_{\rm int}$ is the number of teeth in the integer MFC. For low values of $n$, experimental observations align closely with this linear scaling. However, deviations emerge for larger $n$ (e.g., $n=10$ and $15$), where the observed tooth count falls short of the linear prediction. This discrepancy arises because the power of the weak probe 2 microwave is insufficient to sustain the higher-order nonlinear multi-wave mixing required to fully populate the spectrum, which can be optimized by increasing its power (see SM Sec. II~\cite{supplement}).

To elucidate the physical mechanism underlying the fractional MFC, we consider the nonlinear dynamics of Kittel magnon $\hat{m}_{0}$ with the FMR frequency $\omega_{m}$ in the magnetic sphere.  For pump powers $P_c\le -10$~dBm, the degenerate four-magnon scattering is not triggered because the drive amplitude $\Omega_c$
remains well below the trigger threshold of second-order Suhl instability~\citep{Anderson1955,Suhl1957,Pecora1988,Schultheiss2012,Pirro2014}, as calculated in SM Sec.~III~\cite{supplement}. Consequently, the process $2\omega_m\rightarrow \omega_{\bf q}+\omega_{-{\bf q}}$ does not contribute to the generation of MFC in the current experiments. We thereby construct an effective Hamiltonian that includes the self-Kerr nonlinearity and dipolar parametric three-magnon process. The free Hamiltonian $\hat{H}_{0}/\hbar=\omega_{m} \hat{m}_{0}^{\dagger} \hat{m}_{0}+\omega_{\bf k} \hat{m}_{\bf k}^{\dagger} \hat{m}_{\bf k}+\omega_{-\bf k} \hat{m}_{-\bf k}^{\dagger} \hat{m}_{-\bf k}$
describes the dynamics of Kittel magnon $\hat{m}_{0}$ and a pair of degenerate magnons $\hat{m}_{\pm{\mathbf k}}$ with opposite wave vectors/angular momenta $\pm {\bf k}$~\citep{Ferrimagnetic1959}.  $\hat{H}_{\rm int}/\hbar = g_{\bf k} \hat{m}_{0}^{\dagger} \hat{m}_{\bf k} \hat{m}_{-\bf k}+g_{\bf k}^{*} \hat{m}_{0} \hat{m}_{\bf k}^{\dagger} \hat{m}_{-\bf k}^{\dagger}$ captures the splitting of Kittel magnon into a half-frequency ($\omega_{\bf k}=\omega_{-{\bf k}}=\omega_m/2$) magnon pairs $\hat{m}_{\bf k}$ and $\hat{m}_{-{\bf k}}$ with an amplitude $g_{\bf k}$ via the dipolar interaction~\citep{Three, Saturation,Huang2026}, corresponding to the first-order Suhl instability~\cite{Anderson1955,Suhl1957,Pecora1988}. The Kittel-magnon self-Kerr nonlinearity $\hat{H}_{\rm Kerr}/\hbar = K\hat{m}_{0}^{\dagger} \hat{m}_{0}\hat{m}_{0}^{\dagger} \hat{m}_{0}$ with Kerr coefficient $K$~\citep{Mechanical}, originating from magnetocrystalline anisotropy~\citep{MagnonKerreffect,Zhang2019MagnonKerr,Origin1958,Anisotropy1961}, contributes to the four-wave mixing~\citep{liu2018magnon,2023nonreciprocal}. 
To achieve a fraction-order frequency comb, the YIG sphere is subjected to three distinct microwave fields: the strong pump ($\omega_c$), the probe 1 ($\omega_p$), and the probe 2 ($\omega_f$) couple to the Kittel magnon with the coupling strengths $\Omega_c$, $\Omega_p$, and $\Omega_f$~\citep{microscopic}, described by the Hamiltonian $\hat{H}_{d}/\hbar=i[(\Omega_{c}^{*} e^{i\omega_{c}t}+\Omega_{p}^{*} e^{i\omega_{p}t}+\Omega_{f}^{*} e^{i\omega_{f}t})\hat{m}_{0}-(\Omega_{c} e^{-i\omega_{c}t}+\Omega_{p} e^{-i\omega_{p}t}+\Omega_{f} e^{-i\omega_{f}t})\hat{m}_{0}^{\dagger}]$.

The numerical calculation is performed by the Heisenberg-Langevin equation of motion~\cite{Perina1991}, as detailed in SM Sec.~III~\cite{supplement}.
We find that the contribution from the Kerr effect in the generation of both integer and fractional MFCs is negligible in the YIG sphere (see SM~\cite{supplement} Sec.~III). A detailed comparison between the measurements and calculations by the three-magnon parametric process for various values of $n$ is presented in the End Matter, demonstrating excellent agreement. Notably, at higher orders, certain sum-difference sidebands exhibit amplitudes surpassing adjacent integer orders---a non-perturbative behavior indicating that high-order multi-wave mixing can overcome conventional perturbation constraints~\citep{Optomagnonic,Ghimire2011}.

Further, we perform a perturbation expansion to analytically unveil the three-wave mixing nonlinear cascade mechanism of the fractional MFC. To this end, the driven magnon amplitudes are composed into the driven steady state and their fluctuation.
As derived in SM Sec.~IV~\cite{supplement},  the steady-state scale as $\langle\hat{\alpha}\rangle \propto g_{\mathbf{k}}^{-1}$ and $\langle\hat{\beta}_{\pm\mathbf{k}}\rangle \propto g_{\mathbf{k}}^{-1/2}$. 
By expanding the fluctuations as $\delta\hat{\alpha}(t) = \sum_{N=0}^{\infty} \delta\hat{\alpha}^{(N)}(t)$ and $\delta\hat{\beta}_{\pm \mathbf{k}}(t) = \sum_{N=0}^{\infty} \delta\hat{\beta}_{\pm \mathbf{k}}^{(N)}(t)$, where $\{\delta\hat{\alpha}^{(N)}$, $\delta\hat{\beta}_{\pm{\mathbf k}}^{(N)}\}\propto g_{\bf k}^{N/2} $ for an integer $N\ge 0$, detailed calculation in the SM Sec.~IV~\cite{supplement} reveals that the Kittel-mode fluctuations $\delta\hat{\alpha}^{(N)}$ occur exclusively at \textit{even} orders, while the magnon-pair fluctuations $\delta\hat{\beta}_{\pm\mathbf{k}}^{(N)}$ occur only at \textit{odd} orders,  indicating an alternating driving mechanism between these two modes.

\begin{figure}[htp!]
\begin{center}
    \includegraphics[width=1.02\linewidth]{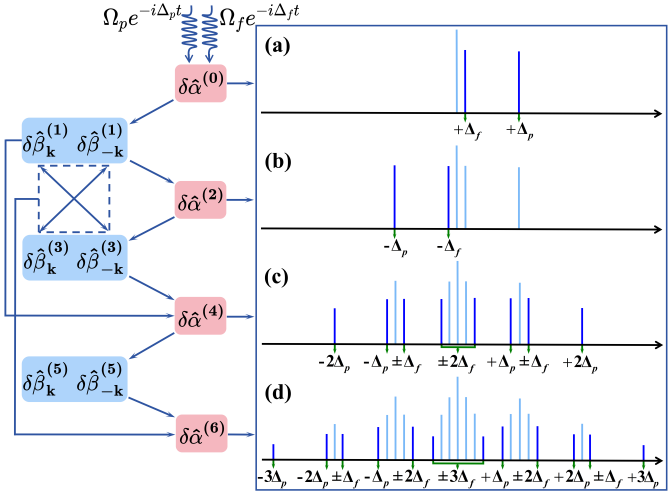}
    \caption{Alternating parity cascade for generating fractional MFC. (a) Zero-order response ($N=0$). Probe 1 and probe 2 microwaves directly excite fluctuations at the positive detuning frequencies $+\Delta_p$ and $+\Delta_f$. (b) Second-order response ($N=2$). Linear feedback from the first-order magnon-pair fluctuations $\delta\hat{\beta}_{\pm\mathbf{k}}^{(1)}$ gives rise to the conjugate sidebands at $-\Delta_p$ and $-\Delta_f$. (c) Fourth-order cascade ($N=4$). Recombination of the first-order magnon-pair fluctuations $\delta\hat{\alpha}^{(4)} \propto \delta\hat{\beta}_{\mathbf{k}}^{(1)}\delta\hat{\beta}_{-\mathbf{k}}^{(1)}$, combined with the feedback from $\delta\hat{\beta}_{\pm\mathbf{k}}^{(3)}$, generates the harmonics $\{\pm 2\Delta_p$, $\pm 2\Delta_f\}$ and the sum-and-difference frequency $\pm\Delta_p \pm \Delta_f$. (d) Sixth-order cascade ($N=6$). Higher-order mixing of lower-order fluctuations produces complex spectral patterns, including harmonics $\pm 3\Delta_p$, $\pm 3\Delta_f$, $\pm\Delta_p \pm 2\Delta_f$, and $\pm2\Delta_p \pm \Delta_f$.}
    \label{Fig3}
\end{center}
\end{figure}

Figure~\ref{Fig3} illustrates the cascade processes governing the frequency comb generation. In the zeroth-order process ($N=0$), the probe microwaves directly excite the Kittel mode, generating the fluctuation $\delta\hat{\alpha}^{(0)}$ with frequency components at $+\Delta_p$ and $+\Delta_f$ [Fig.~\ref{Fig3}(a)].
Driven by $\delta\hat{\alpha}^{(0)}$, the first-order magnon-pair fluctuations $\delta\hat{\beta}_{\pm\mathbf{k}}^{(1)}$ emerge at $\pm\Delta_p$ and $\pm\Delta_f$. 
In the second-order response ($N=2$) [Fig.~\ref{Fig3}(b)], linear feedback from $\delta\hat{\beta}_{\pm\mathbf{k}}^{(1)}$ induces the Kittel fluctuation $\delta\hat{\alpha}^{(2)}$ at $-\Delta_p$ and $-\Delta_f$.
For the third-order cascade process ($N=3$), the third-order magnon fluctuations $\delta\hat{\beta}_{\pm \mathbf{k}}^{(3)}$ arises from both linear feedback and nonlinear three-wave mixing between $\delta\hat{\alpha}^{(0)}$ and  $\delta\hat{\beta}_{\pm\mathbf{k}}^{(1)\dagger}$, generating the integer-order ($\pm 2\Delta_p$, $\pm 2\Delta_f$) and the sum-and-difference frequencies ($\pm\Delta_p \pm \Delta_f$). 
When the cascade order reaches $N=4$ [Fig.~\ref{Fig3}(c)], the sidebands appear in $\delta\hat{\alpha}^{(4)}$ via two coherent interference paths: one is the linear transfer from $\delta\hat{\beta}_{\pm \mathbf{k}}^{(3)}$, and the other is the nonlinear self-mixing $-ig_{\mathbf{k}}\delta\hat{\beta}_{\mathbf{k}}^{(1)}\delta\hat{\beta}_{-\mathbf{k}}^{(1)}$.
Through this alternative order generation mechanism, the nonlinear process evolves continuously to higher orders. As depicted in Fig.~\ref{Fig3}(d), the linear driving from $\delta\hat{\beta}_{\pm\mathbf{k}}^{(5)}$ and the nonlinear recombination of $\delta\hat{\beta}_{\pm\mathbf{k}}^{(1)}$ with $\delta\hat{\beta}_{\pm\mathbf{k}}^{(3)}$ jointly drive a richer set of sidebands of the sixth-order Kittel fluctuation $\delta\hat{\alpha}^{(6)}$, including the frequency components at $\pm 3\Delta_p$, $\pm 3\Delta_f$, $\pm 2\Delta_p \pm \Delta_f$, and $\pm\Delta_p \pm 2\Delta_f$.

\begin{figure}[htb!]
\begin{center}
    \includegraphics[width=1.02\linewidth]{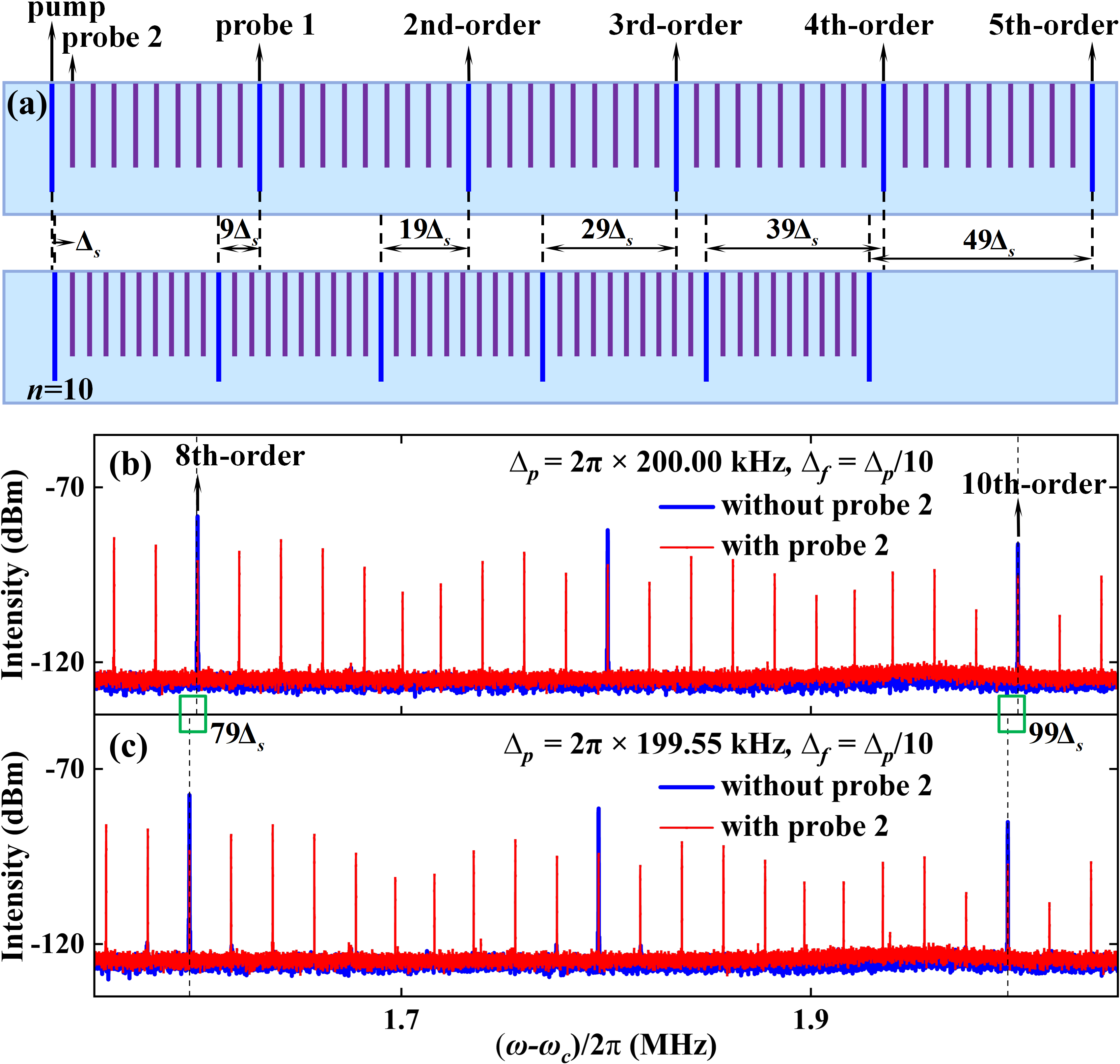}
    \caption{Amplification of frequency shift using the fractional MFC [(a)]. (b) and (c): Experimental verification of frequency-shift amplification using fractional MFC.}
    \label{Fig4}
\end{center}
\end{figure}

We now unlock the capability of fractional MFC for sensitive frequency resolution. Leveraging the fractional-order spectrum structure, we demonstrate sensitive detection of small frequency shifts, as illustrated in Fig.~\ref{Fig4}(a). We introduce a small perturbation $\Delta_s$ to the pump frequency while keeping the probe 2 frequency fixed.
To preserve the fractional MFC structure with $n = 10$, the probe 1 frequency is correspondingly shifted by $9\Delta_s$. Acting as a frequency-domain ``vernier caliper," this small perturbation accumulates systematically across the high-order comb teeth, resulting in a pronounced spectral shift of $(10j-1)\Delta_s$ at the $j$-th integer-order sideband [Fig.~\ref{Fig4}(a)]. By monitoring this amplified frequency shift at a specific high-order comb tooth $j$, we can resolve minute frequency variations with exceptional precision. This is enhanced by a factor of $(n\times j-1)$, effectively surpassing the limitations of conventional measurement techniques.

Figures~\ref{Fig4}(b) and (c) provide a direct experimental verification of the vernier amplification mechanism at fixed powers $\{P_c,P_p,P_f\} = \{-10,-15,-20\}$~dBm for the pump, probe 1, and probe 2 microwaves. Initially,  with the pump fixed at $\omega_c =2\pi\times 3$~GHz, the detunings of the probe 1 and probe 2 relative to the pump are set to $\Delta_p =2\pi\times 0.2$~MHz and $\Delta_f=2\pi\times 0.02$~MHz, generating a fractional MFC with $n=10$ [Fig.~\ref{Fig4}(b)]. We then introduce a minute shift of $\Delta_s=2\pi\times 50$~Hz to the pump frequency, while keeping the probe 2 frequency fixed.  Notably, in this measurement, the spectrum analyzer's resolution bandwidth (RBW) is set to $50$~Hz, making it challenging to directly measure the minute frequency shift of $2\pi\times 50$~Hz in the pump microwave, as it lies at the RBW limit.
However, leveraging the fractional MFC as a vernier caliper, this shift is significantly magnified at high-order teeth. To restore the $n=10$ structure, the frequency of probe 1 drive is shifted by  $-9\Delta_s=-2\pi\times 450$~Hz. Consequently, the 10th-order integer sidebands exhibit amplified shifts of  $-99\Delta_s=-2\pi\times 4950$~Hz. These shifts are far larger than the $50$~Hz RBW and are easily resolvable [Fig.~\ref{Fig4}(c)].

To quantitatively assess the potential of the vernier caliper mechanism for weak magnetic field detection, we consider a scenario where the pump frequency shift $\Delta_s$ is induced by the FMR response to an external field $\Delta B$. According to Kittel's formula, this shift is given by  $\Delta_{s}=\gamma\Delta B$, where $\gamma=2\pi\times 28$~GHz/T is the gyromagnetic ratio. By monitoring the 10th-order integer sideband, the effective magnetic field responsivity is amplified by a factor of 99, yielding $99\Delta_{s}=99\gamma\Delta B=\gamma_{s}\Delta B$, i.e., a giant magnetic field responsivity of up to $\gamma_{s}=99\gamma$ is achieved.  When $\Delta_s$ is within the RBW, the corresponding minimum detectable magnetic field $\Delta B =\Delta_{s}/\gamma$, corresponding to an inferred field sensitivity of  $\sim 1.8$~nT at the RBW limit, illustrates the principle of magnification for weak signal detection. Given that the typical FMR linewidth in YIG is $\sim 2\pi\times 0.3$~MHz, such a resolution is currently unattainable and would require specialized systems with ultranarrow linewidths~\citep{Ultrasensitivejv6l}.

We have experimentally demonstrated a fractional MFC in a high‑quality YIG sphere, identifying the parametric three‑magnon scattering, rather than Kerr nonlinearities or degenerate four-magnon scattering, as the dominant mechanism. By introducing a three-color microwave spectroscopy, we decouple the comb spacing from intrinsic characteristic frequencies, compressing the step to a rational fraction up to $1/20$ and generating high‑density spectral grids exceeding 250 teeth. Unlike conventional four-wave mixing, the three-magnon process induces an alternating parity cascade, where Kittel-mode fluctuations exist only at even orders and magnon-pair fluctuations at odd orders, organizing the spectrum into integer‑order, pure‑fractional, and sum‑difference families. This architecture functions as a frequency‑domain vernier caliper: minute frequency shifts $\Delta_s$ accumulate as $(jn-1)\Delta_s\gg \Delta_s$, becoming resolvable well below the spectrometer's nominal resolution bandwidth. This architecture-independent mechanism is free from constraints of nanofabrication finesse, suggesting a scalable route toward reconfigurable spectral filters, multiplexed sensor arrays, and on‑chip fractional combs.

\begin{acknowledgments}
This work is financially supported by the National Key Research and Development Program of China under Grant No.~2023YFA1406600 and the National Natural Science Foundation of China under Grants No.~12374109 and No.~12274260. G.E.W.B. acknowledges support from JSPS KAKENHI Grants No.~22H04965 and No.~24H02231. 
\end{acknowledgments}

\bibliographystyle{apsrev4-2}
\bibliography{ref}

\section{End Matter}

Disregarding the negligible self-Kerr nonlinearity and accounting for the intrinsic dissipation $\gamma_{0}$ for the Kittel magnon and $\gamma_{\pm \bf k}$ for the magnon pairs, the magnon dynamics are governed by the Heisenberg-Langevin equation of motion~\cite{Perina1991}
\begin{align}
{d \hat{m}_{0}}/{dt}&=-\left(i\omega_{m}+{\gamma_{0}}/{2}\right)\hat{m}_{0}-ig_{\bf k}\hat{m}_{\bf k}\hat{m}_{-\bf k}\nonumber\\
&-\Omega_{c}e^{-i \omega_{c} t}-\Omega_{p}e^{-i \omega_{p} t}-\Omega_{f}e^{-i \omega_{f}t},\nonumber \\
{d \hat{m}_{\pm {\bf k}}}/{dt}&=-\left(i\omega_{\pm {\bf k}}+{\gamma_{\pm {\bf k}}}/{2}\right)\hat{m}_{\pm {\bf k}}-ig_{\bf k}^{*} \hat{m}_{0}\hat{m}_{\mp\bf k}^{\dagger}.
\label{equation_of_motion_a}
\end{align}
We numerically solve Eq.~\eqref{equation_of_motion_a} using the fourth-order Runge-Kutta approach and compute the frequency spectra by applying a fast Fourier transform (FFT) to the time series. For our devices, we choose the parameters according to~\cite{Huang2026}: $\omega_{c}/(2\pi)=\omega_{m}/(2\pi)=3.0000$~GHz, $\omega_{p}/(2\pi)=3.0002$~GHz, $\Delta_{p}/(2\pi)=(\omega_{p}-\omega_{c})/(2\pi)=0.2$~MHz, $P_{c}=-10$~dBm, $P_{p}=-15$~dBm, $P_{f}=-25$~dBm, $\omega_{\bf k}=\omega_{-\bf k}={\omega_c}/2$, $\gamma_0/(2\pi)=1.5$~MHz, $\gamma_{\bf k}/(2\pi)=\gamma_{-\bf k}/(2\pi)=0.11$~MHz, corresponding to the Gilbert damping coefficient $\alpha=0.4\times 10^{-4}$,  and $g_{\bf k}=3$~Hz. In the theoretical calculations, we vary the detuning $\Delta_{f}=\omega_{f}-\omega_{c}=\Delta_{p}/n$ between probe 2 and the pump microwaves to observe the MFC, where $n$ is an integer. We take the first-order anisotropy constant $K_{\rm an}=-610$~J/m$^3$ such that $K\approx -1.58 \times 10^{-10}$~Hz for the YIG sphere of 1~millimeter diameter~\cite{MagnonKerreffect,Zhang2019MagnonKerr,Mechanical} in the calculation.

\begin{figure*}[htp!]
\begin{center}
    \includegraphics[width=0.98\linewidth]{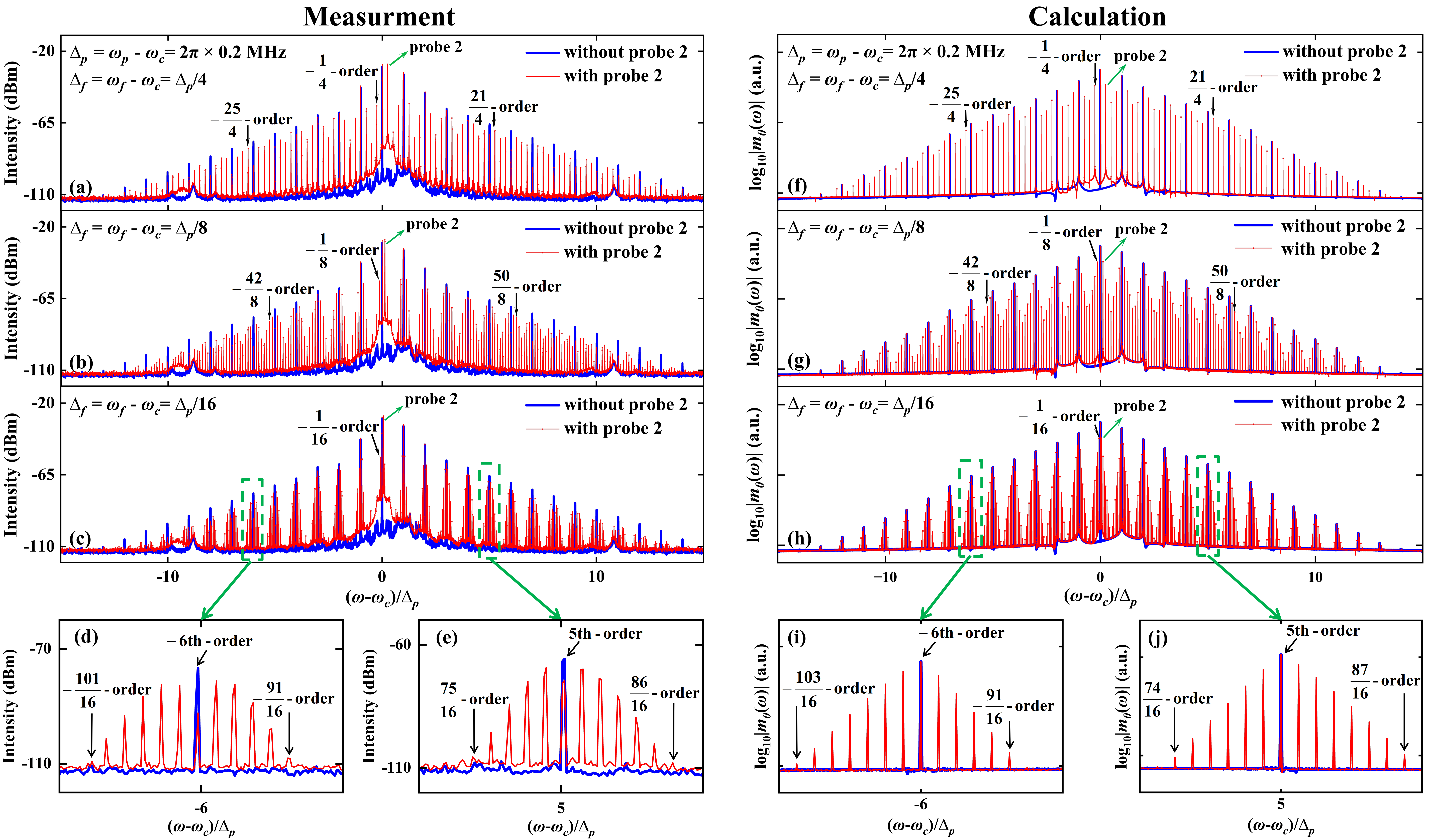}
    \caption{Comparison between measured and calculated spectra of fractional MFC for different fraction control parameters $n$. (a, f), (b, g), and (c, h) present the spectrum comparisons between experiment and theory for  $n = 4$, $8$, and $16$, respectively. (d, e) Magnified experimental spectra of the fractional sideband structures for $n = 16$, highlighting the high-resolution details of the comb teeth. (i, j) Corresponding theoretical magnified spectra for $n = 16$ clearly reproduce the fine fractional structures observed in the experiment. The blue curves depict conventional integer-order combs driven by the pump and probe 1 alone; the red curves denote the fractional MFC generated under the three-microwave driving.}
    \label{Fig5}
\end{center}
\end{figure*}

A detailed comparison between the measurements and calculations for various values of $n$ is presented in Fig.~\ref{Fig5}, demonstrating excellent agreement.
For instance, with the fraction control parameter $n=4$ in Fig.~\ref{Fig5}(a) and (f), the spectral analysis reveals exactly three equidistant fractional-order teeth between adjacent integer-order sidebands, effectively reducing the spacing to one-quarter of the original interval. Our model calculation in Fig.~\ref{Fig5}(f) accurately captures the spectral distribution described by $\Omega_{\rm III}=\omega_c\pm (j\pm l/n)\Delta_p$, demonstrating that the fractional MFC originates from multi-wave mixing driven by the three-magnon interaction.

As $n$ increases in Fig.~\ref{Fig5}(b) and (c), the integer-order sidebands in $\Omega_{\rm I}=\omega_c \pm j\Delta_p$ remain prominent, and their positions are completely unaffected by $n$. For instance, when $n=8$, the frequency spacing is reduced to $\Delta_f=\Delta_p/8$, and the high-order multi-wave mixing governed by the three-magnon scattering generates seven fractional-order comb teeth within each neighboring integer-order interval. 
Comparing Fig.~\ref{Fig5}(b) and (g), the theoretical model accurately reproduces the fine structure of the fractional MFC, including the marked $-1/8$-order, $-42/8$-order, and $50/8$-order sidebands. 
As $n$ increases to $16$, Figs.~\ref{Fig5}(c) and (h) show the continuous emergence of sidebands at intervals of $\Delta_p/16$ surrounding the integer orders. To directly compare experiment and theory, spectral regions near the $-6$th- and $5$th-orders are extracted. 
Figure~\ref{Fig5}(d) and (e) display the magnified views of experimental observations, while Figs.~\ref{Fig5}(i) and (j) show the corresponding theoretical calculations. 
As shown in Fig.~\ref{Fig5}(d), in the negative frequency region near the $-6$th-order integer-order sideband, the experimentally observed fractional-order comb teeth exhibit a cutoff-order precisely at the $-101/16$-order and the $-91/16$-order, while the theoretical calculation in Fig.~\ref{Fig5}(i) perfectly reproduces this cutoff-order. In the positive frequency region near the $5$th-order integer sideband in Fig.~\ref{Fig5}(e), the measured spectra exhibit a symmetric structure with the fractional-order sidebands distributed between the $75/16$-order and the $86/16$-order, which is well reproduced by the theoretical calculations in Fig.~\ref{Fig5}(j).

The theoretical results in Figs.~\ref{Fig5}(i) and (j) indicate that the amplitudes of the fractional-order ($\Omega_{\rm II}$) and sum-difference frequency sidebands ($\Omega_{\rm III}$) should remain below those of the adjacent integer-order sidebands. This is because the primary energy transfer pathway favors the integer orders, while the fractional orders arise from secondary, less efficient mixing processes. However, the experimental measurements in Figs.~\ref{Fig5}(d) and (e) reveal an anomalous enhancement, where certain sum-difference frequency sidebands exhibit amplitudes surpassing those of the adjacent integer orders. This non-perturbative behavior suggests that high-quality YIG spheres driven by multi-frequency microwaves can overcome the limitations of conventional perturbation theory, enabling exceptionally high-order nonlinear frequency conversion~\citep{Optomagnonic,Ghimire2011}.

\end{document}